\documentclass{PoS}

\usepackage{graphicx}
\usepackage{amsmath}
\usepackage{amssymb}

\newcommand{\be}{\begin{equation}}
\newcommand{\ee}{\end{equation}}
\newcommand{\bea}{\begin{eqnarray}}
\newcommand{\eea}{\end{eqnarray}}
\newcommand{\bi}{\begin{itemize}}
\newcommand{\ei}{\end{itemize}}
\newcommand{\ben}{\begin{enumerate}}
\newcommand{\een}{\end{enumerate}}
\newcommand{\bt}{\begin{tabbing}}
\newcommand{\et}{\end{tabbing}}
\newcommand{\nn}{\nonumber}

\newcommand{\calO}{{\mathcal O}}

\newcommand{\crad}{\langle r^2 \rangle}
\newcommand{\pp}{{p^\prime}}
\newcommand{\bfp}{{\bf p}}
\newcommand{\bfpp}{{{\bf p}^\prime}}

\newcommand{\bfx}{{\bf x}}
\newcommand{\bfxp}{{{\bf x}^\prime}}
\newcommand{\bfxpp}{{{\bf x}^{\prime\prime}}}
\newcommand{\bfr}{{\bf r}}
\newcommand{\bfz}{{\bf 0}}
\newcommand{\dt}{{\Delta t}}
\newcommand{\dtp}{{\Delta t^\prime}}

\title{
   \begin{picture}(0,0)(0,0)%
   \put(350,75){\makebox(0,0)[l]{\textnormal{\normalsize KEK-CP-278}}}%
   \put(350,60){\makebox(0,0)[l]{\textnormal{\normalsize OU-HET-765-2012}}}%
   \put(350,45){\makebox(0,0)[l]{\textnormal{\normalsize UTHEP-646}}}%
   \end{picture}%
   Chiral behavior of kaon semileptonic form factors 
   in lattice QCD with exact chiral symmetry
}

\ShortTitle{
   Chiral behavior of kaon semileptonic form factors 
   in lattice QCD with exact chiral symmetry
}

\author{
   JLQCD Collaboration: 
   \speaker{T.~Kaneko}$^{a,b}$\thanks{E-mail: takashi.kaneko@kek.jp}, 
   S.~Aoki$^{c,d}$, 
   G.~Cossu$^a$, 
   X.~Feng$^a$, 
   H.~Fukaya$^e$, 
   S.~Hashimoto$^{a,b}$, 
   J.~Noaki$^{a}$
   and
   T.~Onogi$^e$
   \\
   \\
   \\
   \llap{$^a$}
   High Energy Accelerator Research Organization (KEK),
   Ibaraki 305-0801, Japan 
   \\
   \llap{$^b$}
   School of High Energy Accelerator Science,
   The Graduate University for Advanced Studies (Sokendai),
   Ibaraki 305-0801, Japan
   \\ 
   \llap{$^c$}
   Graduate School of Pure and Applied Sciences, 
   University of Tsukuba, Ibaraki 305-8571, Japan
   \\ 
   \llap{$^d$}
   Center for Computational Sciences, University of Tsukuba, Tsukuba, 
   Ibaraki 305-8577, Japan
   \\
   \llap{$^e$}
   Department of Physics, Osaka University, 
   Toyonaka, Osaka 560-0043, Japan
}

\abstract{
We calculate the kaon semileptonic form factors in lattice QCD 
with three flavors of dynamical overlap quarks. Gauge ensembles 
are generated at pion masses as low as 290 MeV and at a strange 
quark mass near its physical value. We precisely calculate 
relevant meson correlators using the all-to-all quark propagator.
Twisted boundary conditions and the reweighting technique are 
employed to vary the momentum transfer and the strange quark mass. 
We discuss the chiral behavior of the form factors by comparing 
with chiral perturbation theory and experiments.
}

\FullConference{
   The 30 International Symposium on Lattice Field Theory - Lattice 2012,\\
   June 24-29, 2012\\
   Cairns, Australia
}

\begin{document}


\section{Introduction}


Precise determination of the Cabibbo-Kobayashi-Maskawa (CKM) matrix elements
provides a stringent test of the Standard Model to search for new physics. 
One of the elements $|V_{us}|$ can be determined 
from the $K \!\to\! \pi l \nu$ decays
through a theoretical calculation of the normalization
of the vector form factor $f_+(0)$, which parametrizes 
the $K \to \pi$ matrix element
\bea
   \langle \pi(p^\prime) | V_\mu | K(p) \rangle 
   & = & 
   (p+p^\prime)_\mu f_+(q^2) + (p-p^\prime)_\mu f_-(q^2)
   \hspace{3mm}
   (q^2=(p-p^\prime)^2).
   \label{eqn:intro:ME}
\eea
In this article,
we report on our calculation of $f_+(0)$ in $N_f\!=\!2\!+\!1$ lattice QCD
within 1\,\% accuracy.
In order to demonstrate the reliability of this precise calculation,
we also examine the consistency of other quantities,
such as $f_-(0)$ and the form factors' shape, 
with chiral perturbation theory (ChPT) and experiments.


\section{Calculation of form factors}


We employ the overlap quark action to exactly preserve chiral symmetry
for a straightforward comparison of $f_{\{+,-\}}(q^2)$ with ChPT.
Numerical simulations are accelerated 
by modifying the Iwasaki gauge action~\cite{exW}
and by simulating the trivial topological sector~\cite{exW,fixedQ}.
Effects of the fixed topology are suppressed by the inverse of the 
lattice volume $N_s^3 \times N_t$~\cite{fixedQ}.
We use our gauge ensembles generated at a single lattice spacing 
$a\!=\!0.112(1)$~fm,
which is determined from the $\Omega$ baryon mass,
and at a strange quark mass $m_s\!=\!0.080$,
which is close to its physical value $m_{s,\rm phys}\!=\!0.081$.
We simulate four values of degenerate up and down quark masses 
$m_{ud}\!=\!0.015$, 0.025, 0.035 and 0.050
that cover a range of the pion mass 290\,--\,540~MeV.
At each $m_{ud}$, 
a lattice size of $16^3 \!\times\! 48$ or $24^3 \!\times\! 48$ 
is chosen to control finite volume effects
by satisfying a condition $M_\pi L \! \gtrsim \! 4$.
The statistics are 2,500 HMC trajectories 
at each combination of $m_{ud}$ and $m_s$.


We calculate the scalar form factor 
$f_0(q^2)\!=\!f_+(q^2)+f_-(q^2)\,q^2/(M_K^2-M_\pi^2)$
at $q^2\!=\!q^2_{\rm max}\!=\!(M_K-M_\pi)^2$,
$f_+(q^2)$ and $\xi(q^2)\!=\!f_-(q^2)/f_+(q^2)$ at $q^2 < q^2_{\rm max}$ 
from the following ratios~\cite{kl3:drat}
\bea
   R
   & = & 
   \frac{C^{K \pi}_4(\dt,\dtp; \bfz, \bfz)
         C^{\pi K}_4(\dt,\dtp; \bfz, \bfz)}
        {C^{K K}_4(\dt,\dtp; \bfz, \bfz)
         C^{\pi \pi}_4(\dt,\dtp; \bfz, \bfz)}
   \xrightarrow[\dt, \dtp \to \infty]{} 
   \frac{(M_K+M_\pi)^2}{4 M_K M_\pi} f_0(q^2_{\rm max})^2,
   \label{eqn:drat:drat1}
   \\[1mm]
   \tilde{R} 
   & = & 
   \frac{C^{K \pi}_4(\dt,\dtp; \bfp, \bfpp)
         C^{K}(\dt, \bfz)\, C^{\pi}(\dtp, \bfz)}
        {C^{K \pi}_4(\dt, \dtp; \bfz, \bfz)
         C^{K}(\dt,\bfp)\, C^{\pi}(\dtp, \bfpp)}
   %
   \to 
   \left\{
      \frac{E_K+E_\pi^\prime}{M_K+M_\pi} 
    + \frac{E_K-E_\pi^\prime}{M_K+M_\pi} \xi(q^2)
   \right\}
   \frac{f_+(q^2)}{f_0(q^2_{\rm max})},
   \hspace{9mm}
   \label{eqn:drat:drat2}
   \\[1mm]
   R_k 
   & = & 
   \frac{C^{K \pi}_k(\dt,\dtp; \bfp, \bfpp)
         C^{KK}_4(\dt, \dtp; \bfp, \bfpp)}
        {C^{K \pi}_4(\dt, \dtp; \bfp, \bfpp)
         C^{KK}_k(\dt, \dtp; \bfp, \bfpp)}
   \to 
   \frac{E_K+E_K^\prime}{(p+\pp)_k}
   \frac{(p+\pp)_k + (p-\pp)_k \xi(q^2)}
        {E_K+E_\pi^\prime + (E_K-E_\pi^\prime)\xi(q^2)}
   \label{eqn:drat:drat3}
\eea
where 
$E_P^{(\prime)}$ ($P=\pi$ or $K$) 
represents the meson energy with the spatial momentum $\bfp^{(\prime)}$.
Note that these observables are sufficient to determine
$f_{\{+,-,0\}}(q^2)$ at simulated values of $q^2$,
except at $q^2_{\rm max}$,
where $\tilde{R}$ and $R_k$ have no sensitivity to $\xi(q^2)$.
Two- and three-point functions are defined as 
\bea
   C^P(\dt,\bfp)
   & = &
   \frac{1}{N_s^3 N_t}\sum_{\bfx,t} 
   \sum_{\bfxp}
   \langle 
      \calO_P(\bfxp,t+\dt) \calO_P^\dagger(\bfx,t) 
   \rangle
   e^{-i \bfp (\bfxp-\bfx)},
   \label{eqn:sim_param:msn_corr_2pt}
   \\
   C_\mu^{PQ}(\dt,\dtp;\bfp,\bfpp)
   & = &
   \frac{1}{N_s^3 N_t}\sum_{\bfx,t} 
   \sum_{\bfxpp, \bfxp}
   \langle 
      \calO_Q(\bfxpp,t+\dt+\dtp) V_\mu(\bfxp,t+\dt) \calO_P^\dagger(\bfx,t)
   \rangle
   \nn
   \\
   && 
   \hspace{54mm} \times e^{-i \bfpp (\bfxpp-\bfxp)-i \bfp (\bfxp-\bfx)},
   \label{eqn:sim_param:msn_corr_3pt}
\eea
where $\calO_{P(Q)}^\dagger(\bfx,t)$ ($P,Q = \pi$ or $K$) 
represents 
the meson interpolating operator 
$\sum_{\bfr} \phi(\bfr) \bar{q}(\bfx+\bfr,t) \gamma_5 q^\prime(\bfx,t)$
with an exponential smearing function $\phi(\bfr)\!=\!e^{-0.4|\bfr|}$.
We use the all-to-all propagator~\cite{A2A,prev0}
to remarkably improve the statistical accuracy of the correlators
and, hence, form factors~\cite{prev1,prev2}.
%


In order to precisely determine $f_+(0)$, 
we explore the most important kinematical region $q^2\!\sim\!0$
by using the twisted boundary conditions (TBCs) \cite{TBC}
\bea
   q(\bfx+N_s\,\hat{k},t) = e^{i\theta}q(\bfx,t),
   \hspace{3mm}
   \bar{q}(\bfx+N_s\,\hat{k},t) = e^{-i\theta}\bar{q}(\bfx,t)
   \hspace{3mm} 
   (k=1,2,3),
   \label{eqn:intro:TBC}
\eea
where $\hat{k}$ is a unit vector in the $k$-th direction.
For simplicity,
we use a common twist angle $\theta$ in all three spatial directions
and take four (three) values of $\theta$ (including $\theta\!=\!0$)
to cover $-0.1\!\lesssim\!q^2[\mbox{GeV}^2]\!\leq\! q^2_{\rm max}$
on the $16^3\!\times\!48$ ($24^3\!\times\!48$) lattice.


\begin{figure}[t]
\begin{center}
   \includegraphics[angle=0,width=0.47\linewidth,clip]%
                   {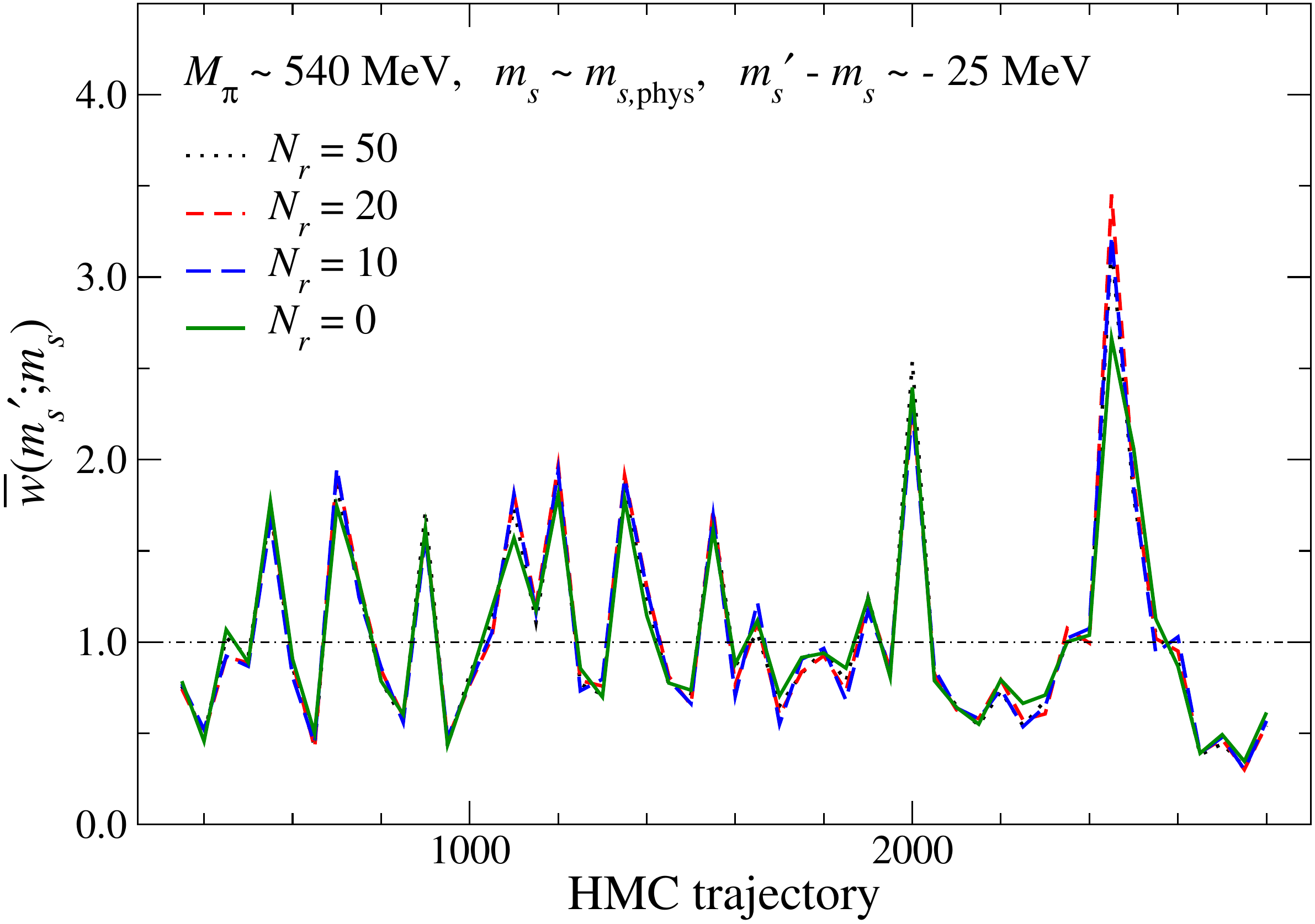}
   \hspace{3mm}
   \includegraphics[angle=0,width=0.49\linewidth,clip]%
                   {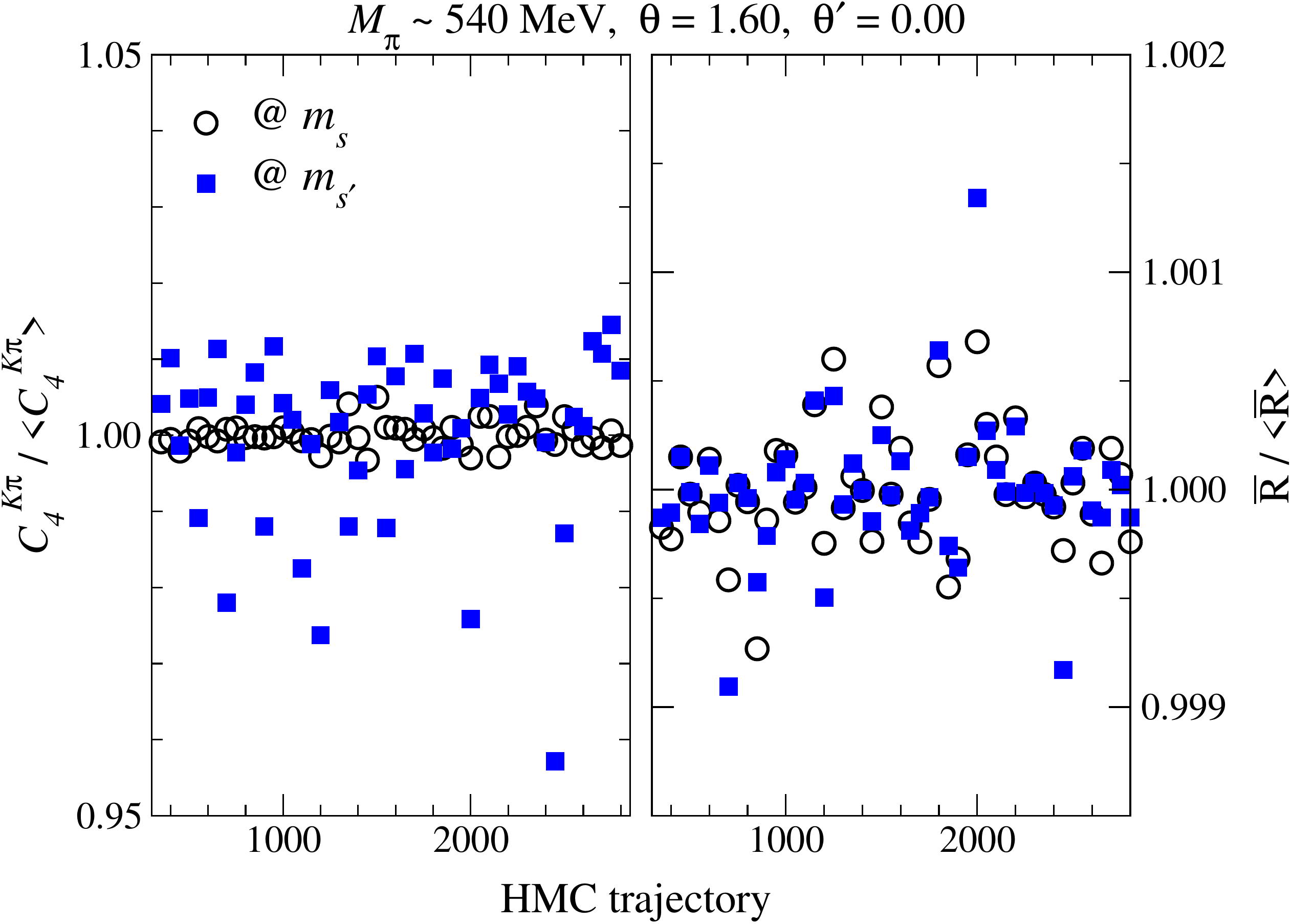}
   \vspace{-1mm}
   \caption{
     Left panel: 
     Monte Carlo history of normalized reweighting factor 
     $\tilde{w}(m_s^\prime,m_s)$ at our largest $m_{ud}$. 
     Four lines show data with different numbers of noise samples ($N_r$).
     Right panel: 
     statistical fluctuation of $C^{K\pi}_4$ and $\tilde{R}$ 
     before (open symbols) and after (filled symbols) reweighting.
     $\theta^{(\prime)}$ represents the twist angle used to induce 
     the initial (final) meson momentum $\bfp^{(\prime)}$.
   }
   \label{fig:reweight:reweight}
\end{center}
\vspace{-4mm}
\end{figure}

\section{Reweighting}

The strange quark mass dependence of the form factors
is studied by simulating a different value $m_s^\prime\!=\!0.060$,
which is about 25\,MeV smaller than $m_s\!=\!0.080$. 
We employ the reweighting technique in which 
an observable $\calO$ at $m_s^\prime$ 
is calculated on the gauge configurations at $m_s$ 
as 
\bea
\langle \calO \rangle_{m_s^\prime}
=
\langle \calO \, \tilde{w}(m_s^\prime,m_s) \rangle_{m_s},
\hspace{3mm}
\tilde{w}(m_s^\prime,m_s)
= 
\frac{w(m_s^\prime,m_s)}{\langle w(m_s^\prime,m_s) \rangle_{m_s}},
\hspace{3mm}
w(m_s^\prime,m_s)
= 
\frac{\det[D(m_s^\prime)]}{\det[D(m_s)]},
\label{eqn:reweight:reweight}
\eea
where 
$\langle \cdots \rangle_{m_s^{(\prime)}}$ and $D(m_s^{(\prime)})$ 
represent the Monte Carlo average and the overlap-Dirac operator 
at the strange quark mass $m_s^{(\prime)}$, respectively.
We consider decomposing the reweighting factor $w$ 
into contributions of low- and high-modes of $D$ 
as $w = w_{\rm low}\, w_{\rm high}$.
The low-mode contribution 
$w_{\rm low} = \prod_k \lambda_k(m_s^\prime) / \prod_k \lambda_k(m_s)$
is exactly calculated by using 160 (240) low-lying eigenvalues $\lambda_k$
on $16^3 \! \times \! 48$ ($24^3 \! \times \! 48$). 
A noisy estimator is employed for 
$w_{\rm high}^2
 = 
 (1/N_r) \sum_{r=1}^{N_r} \exp[ -(\bar{P} \xi_r)^\dagger (\Omega-1) (\bar{P} \xi_r) / 2]$
with 
$\Omega\!=\!D(m_s)^\dagger \left\{ D(m_s^\prime)^{-1}\right\}^\dagger
            D(m_s^\prime)^{-1} D(m_s)$, 
where 
$\{\xi_1,\ldots,\xi_{N_r}\}$ is a set of Gaussian noise vectors, 
and $\bar{P}$ projects them to the eigenspace spanned by the high-modes. 

The left panel of Fig.~\ref{fig:reweight:reweight}
shows the Monte Carlo history of the normalized reweighting factor 
$\tilde{w}$, which appears in the expression of 
the observable in Eq.~(\ref{eqn:reweight:reweight}).
With our simulation setup, 
$\tilde{w}$ has small dependence on the number of the noise vectors $N_r$.
From this observation,
we set $N_r\!=\!10$ to precisely calculate $\tilde{w}$
on each gauge configuration.


The right panel of Fig.~\ref{fig:reweight:reweight}
shows how reweighting affects the statistical accuracy 
of the correlator $C^{K\pi}_4$ and the ratio $\tilde{R}$.
The fluctuation of $C^{K\pi}_4$ is largely enhanced
by $\tilde{w}$, which is typically in the range of $[0.5,2.0]$. 
We observe, however, that 
the enhanced fluctuations are largely canceled in the ratio $\tilde{R}$.
Consequently,
the statistical accuracies of the form factors 
are not largely impaired by reweighting: 
typically $\lesssim$\,1.0\,\% (20\,\%) for $f_{\{+,0\}}(q^2)$ $\left(\xi(q^2)\right)$
before reweighting,
and $\lesssim$\,1.5\,\% (30\,\%) after reweighting.



\begin{figure}[b]
\vspace{2mm}
\begin{center}
   \includegraphics[angle=0,width=0.513\linewidth,clip]%
                   {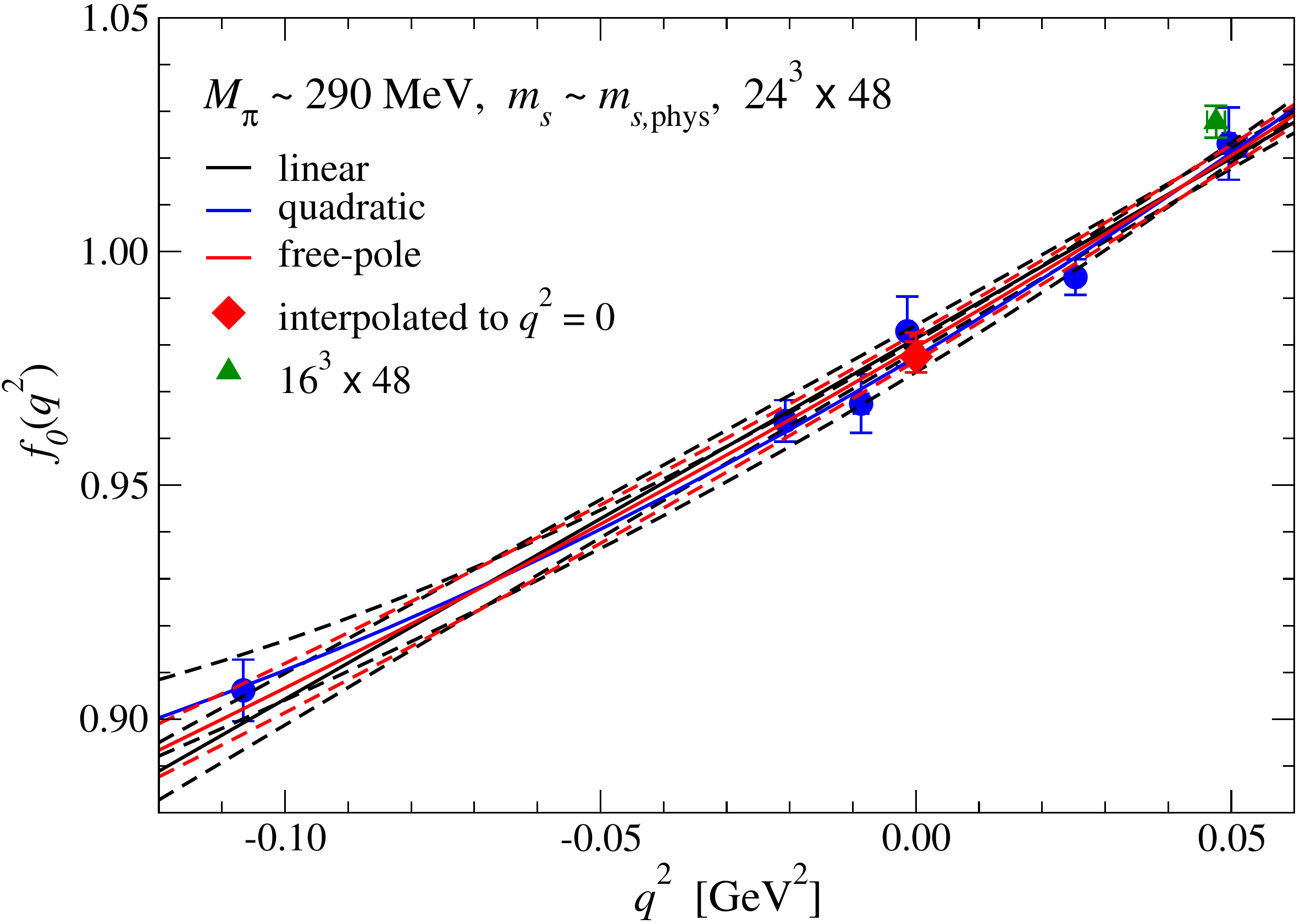}
   \hspace{3mm}
   \includegraphics[angle=0,width=0.453\linewidth,clip]%
                   {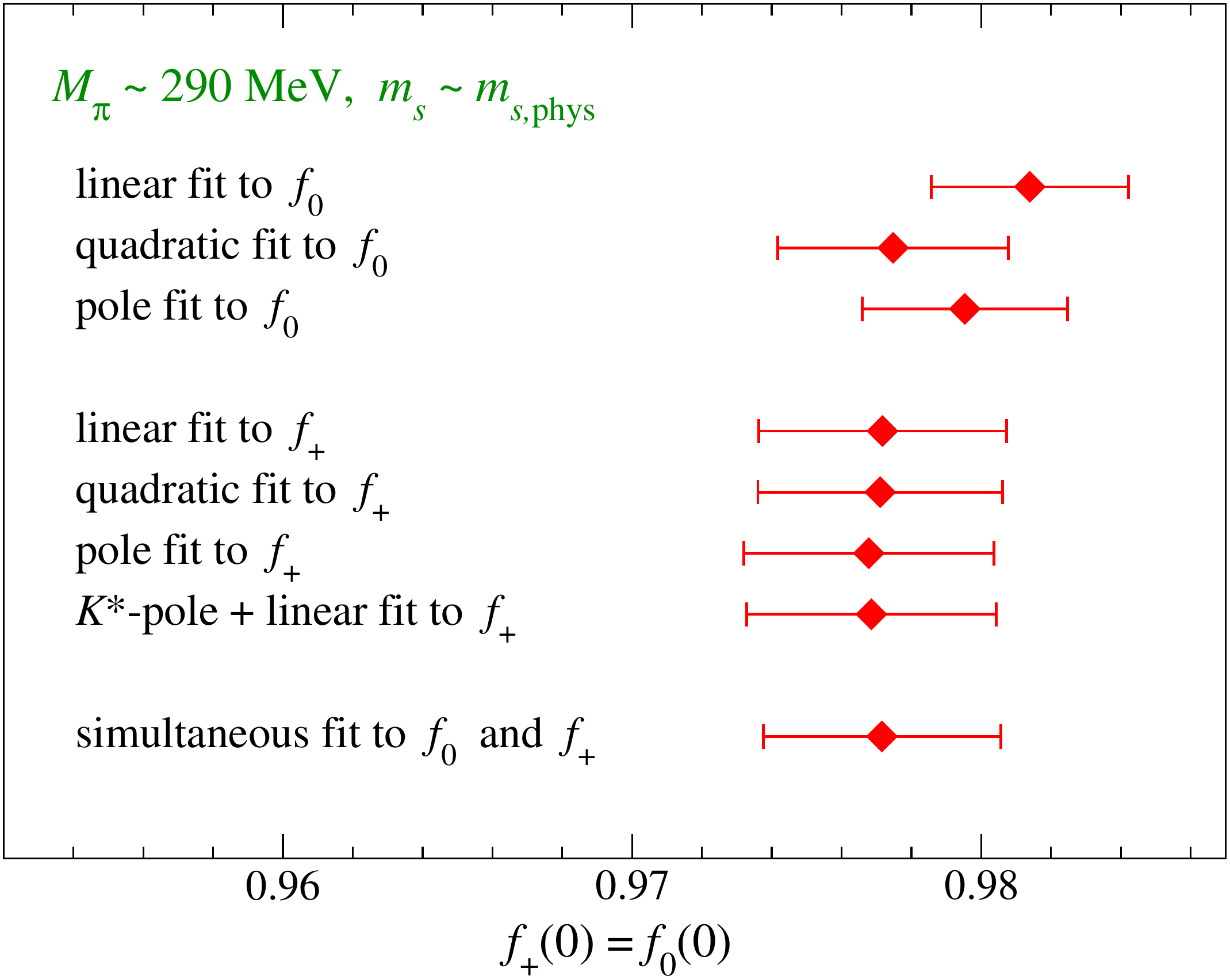}
   \vspace{-5mm}
   \caption{
      Left panel: 
      scalar form factor $f_0(q^2)$ at our smallest $m_{ud}$ 
      as a function of $q^2$. 
      We plot interpolations listed in Eq.~(\protect\ref{eqn:q2-dep:f0})
      together with $f_0(0)$ determined from the quadratic parametrization
      (diamond).
      Our result on a smaller lattice $16^3 \! \times \! 48$ 
      is also plotted to examine finite volume effects (triangle).
      Right panel:
      comparison of $f_+(0)$ $\left(=\!f_0(0)\right)$ obtained by 
      using different parametrization forms.
   }
   \label{fig:q2_dep:f0}
\end{center}
\end{figure}

\section{$q^2$ dependence of form factors}

Our results for $f_0(q^2)$ at the smallest $m_{ud}$ 
are plotted as a function of $q^2$ 
in the left panel of Fig.~\ref{fig:q2_dep:f0}.
We use TBCs to simulate small values of $|q^2|$,
where contributions of higher orders in $q^2$ are small
and our data are well described 
by any of the following parametrization forms
\bea
   f_0(q^2) 
   = 
   f_0(0)(1+c_1 q^2),
   \hspace{5mm}
   f_0(q^2) 
   = 
   f_0(0)(1+c_1 q^2+c_2 q^4), 
   \hspace{5mm}
   f_0(q^2) 
   = 
   \frac{f_0(0)}{1-q^2/M_{\rm pole}^2}.
   \hspace{2mm}
   \label{eqn:q2-dep:f0}
\eea
A different form 
$f_+(0)\left\{1/(1-q^2/M_{K^*}^2)+c_1 q^2\right\}$
with the $K^*$ pole plus a polynomial correction
is also tested for $f_+(q^2)$.
We confirm a good agreement among $f_+(0)$ and $f_0(0)$ 
obtained from these parametrizations
as shown in the right panel of Fig.~\ref{fig:q2_dep:f0}.
In this report,
we employ a simultaneous fit using the quadratic form for $f_0$
and the form with the $K^*$ pole for $f_+$
to determine the normalization $f_+(0)$ and its slope $f^\prime_+(0)$.
The uncertainty of this interpolation 
is estimated by using the different parametrization forms
and turns out to be similar to or smaller than the statistical error.

In the left panel of Fig.~\ref{fig:q2_dep:f0},
we also plot a result of $f_0(q^2)$ 
obtained on a smaller lattice $16^3 \! \times \! 48$.
The difference from $24^3 \! \times \! 48$ can be attributed 
to the conventional finite volume effect as well as the fixed topology effect,
but turns out to be insignificant (0.8\,\%, 2.1\,$\sigma$).
The finite volume effect remaining on the larger volume $24^3 \! \times \! 48$
is estimated as $\lesssim 0.3$\,\% by assuming the $1/N_s^3 N_t$ scaling of the 
fixed topology effect and can be safely neglected in the following analysis.


\section{Chiral extrapolation of $f_+(0)$}

There are two popular choices of the expansion parameter in ChPT:
$\xi_P\!=\!M_P^2/(4 \pi F_\pi)^2$ and 
$M_P^2/(4 \pi F_0)^2$ ($P\!=\!\pi,K,\eta$),
where $F_0$ is the decay constant in the $SU(3)$ chiral limit.
The latter choice largely enhances 
the chiral corrections in $f_+(0)$~\cite{prev2}
and leads to worse convergences in the expansions of 
$M_P$ and $f_P$~\cite{Spectrum:JLQCD}.
We therefore use the former 
and denote the chiral expansion as $f_+(0)\!=\!1+f_2+\Delta f$,
where $f_2$ and $\Delta f$ are $O(\xi_P)$ 
and higher order contributions, respectively.

\begin{figure}[b]
\vspace{2mm}
\begin{center}
   \includegraphics[angle=0,width=0.484\linewidth,clip]%
                   {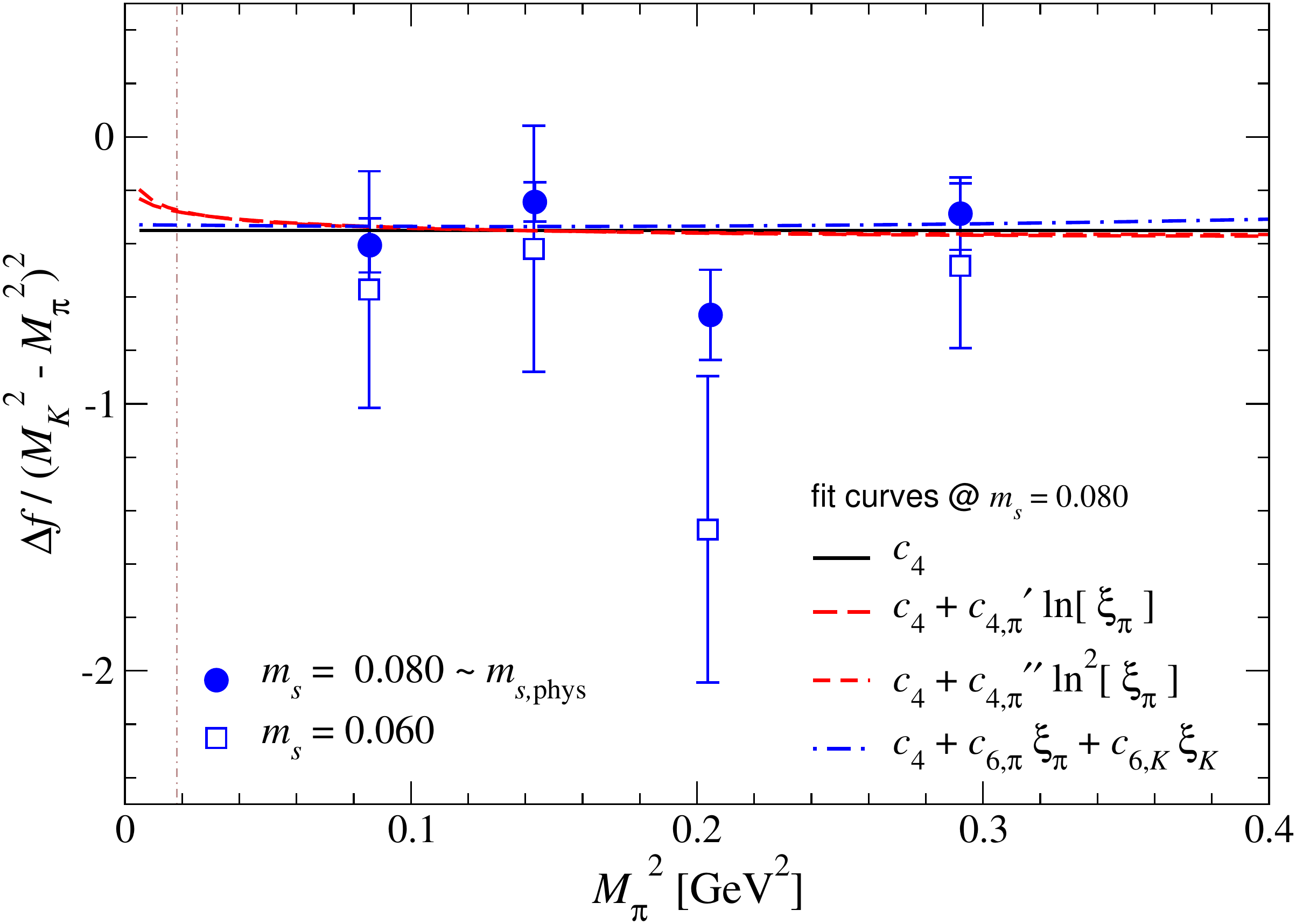}
   \hspace{3mm}
   \includegraphics[angle=0,width=0.484\linewidth,clip]%
                   {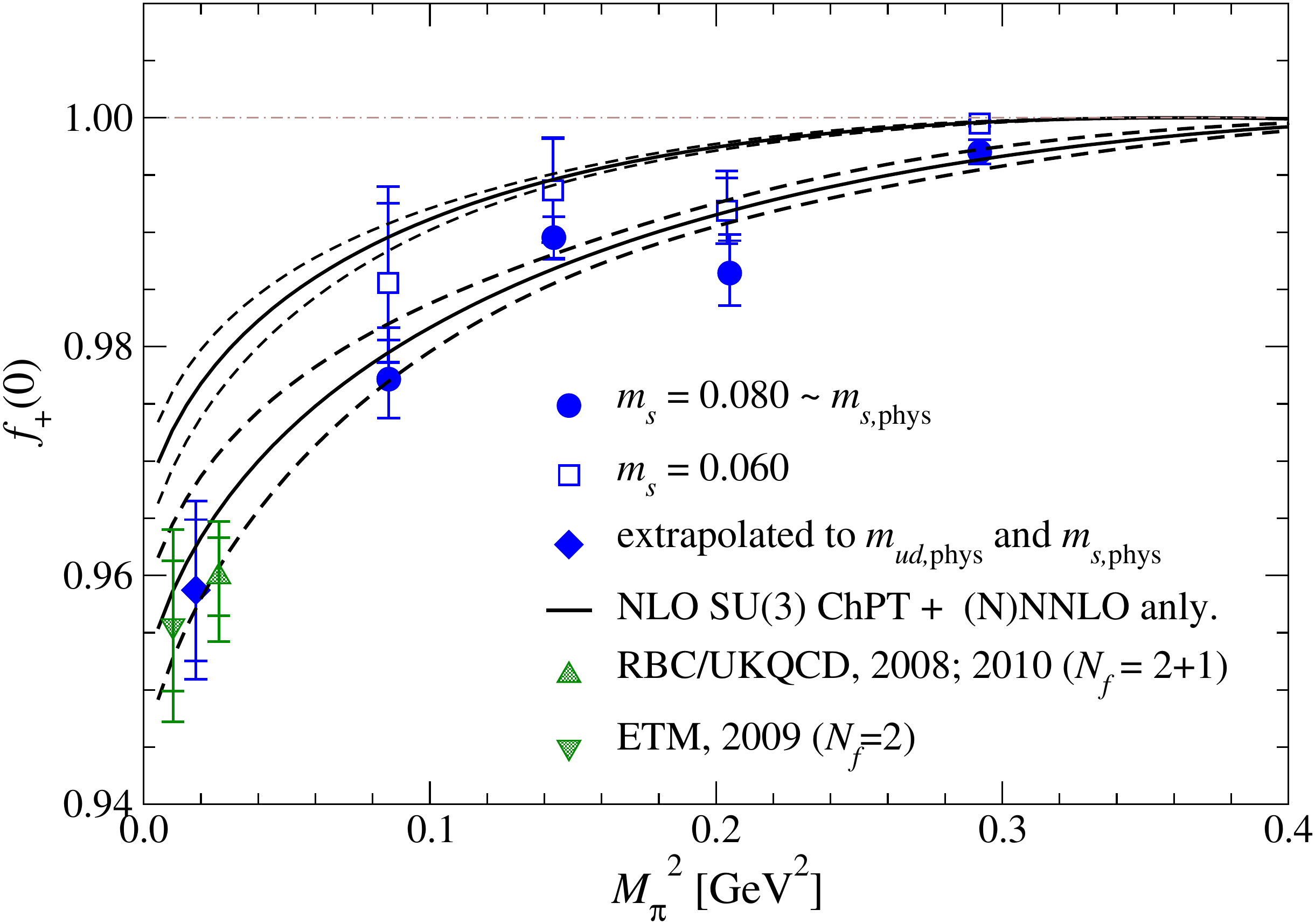}
   \caption{
     Chiral extrapolations of $\Delta f/(M_K^2-M_\pi^2)^2$ (left panel)
     and $f_+(0)$ (right panel). 
     Circles and squares show data at different values of $m_s$.
     In the right panel,
     we also plot $f_+(0)$ from recent calculations
     in $N_f\!=\!2+1$~\cite{kl3:Nf3:RBC/UKQCD} 
     and $N_f\!=\!2$~\cite{kl3:Nf2:ETM} QCD.
   }
   \label{fig:chiral_fit:f+0:f+0}
\end{center}
\vspace{-5mm}
\end{figure}

The Ademollo-Gatto theorem
$f_+(0) -1 \!\propto\! (m_s-m_{ud})^2$~\cite{kl3:AG_theorem} 
guarantees that $f_2$~\cite{kl3:ChPT:f2}
\bea
   f_2 
   = 
   \frac{3}{2} \left( H_{K\pi} + H_{K\eta} \right),
   \hspace{3mm}
   H_{PQ} 
   = 
   -\frac{\xi_P+\xi_Q}{8}
   \left( 1 + \frac{2 \xi_P \xi_Q}{\xi_P^2-\xi_Q^2}
          \ln\left[ \frac{\xi_Q}{\xi_P} \right] \right)
\eea
is written in terms of physical observables $\xi_{\{\pi,K,\eta\}}$
and does not contain low-energy constants (LECs) in the ChPT Lagrangian.
The chiral expansion of $f_+(0)$ is then nothing but the 
parametrization of $\Delta f$. 
In the left panel of Fig.~\ref{fig:chiral_fit:f+0:f+0},
we examine the quark mass dependence of $\Delta f$
divided by $(M_K^2-M_\pi^2)^2$ which is motivated by the Ademollo-Gatto
theorem.
The mild dependence suggests that 
our data 
can be described by a simple constant fit $\Delta f/(M_K^2-M_\pi^2)^2 = c_4$
which gives rise to the $O(\xi_P^2)$ analytic term in $f_+(0)$.
We also confirm that 
the chiral extrapolation of $\Delta f/(M_K^2-M_\pi^2)^2$ is not largely modified
by including the following 
$O(\xi_P^2)$ logarithmic and $O(\xi_P^3)$ analytic corrections 
\bea
   \Delta f / (M_K^2 - M_\pi^2)^2 - c_4 
   \hspace{2mm} = \hspace{1mm}
   c_{4,\pi}^\prime \log[\xi_\pi],
   \hspace{2mm} \mbox{ or } \hspace{1mm}
   c_{4,\pi}^{\prime\prime} \log^2[\xi_\pi],
   \hspace{2mm} \mbox{ or } \hspace{1mm}
   c_{6,\pi} \xi_\pi,
   \hspace{2mm} \mbox{ or } \hspace{1mm}
   c_{6,\pi} \xi_\pi + c_{6,K} \xi_K.
   \hspace{4mm}
\eea

In this report, 
we employ a parametrization 
$\Delta f/(M_K^2-M_\pi^2) = c_4 + c_{6,\pi} \xi_\pi$,
all parameters of which are determined reasonably well
with an acceptable value of $\chi^2/{\rm d.o.f.} \! \sim \! 1.6$.
The systematic uncertainty of this extrapolation 
is estimated as the largest deviation in $f_+(0)$ 
among the different parametrization forms discussed above. 
The discretization error in the $SU(3)$ breaking effect 
$f_2\!+\!\Delta f$ is estimated 
by an order counting $O((a\Lambda_{\rm QCD})^2)$ 
with $\Lambda_{\rm QCD}\!\approx\!500$.
We then obtain 
\bea
   f_+(0) 
   = 
   0.959(6)_{\rm stat}(4)_{\rm chiral}(3)_{a \ne 0},
   \hspace{3mm}
   |V_{us}| 
   = 
   0.2256(19)_{\rm theory+exp't},
\eea
where we use $|V_{us}f_+(0)|\!=\!0.2163(5)$
determined from the $K \!\to\! \pi l \nu$ decay rates~\cite{FlaviaNet:Vusf+0}.
Note that previous calculations in $N_f\!=\!2+1$~\cite{kl3:Nf3:RBC/UKQCD} 
and $N_f\!=\!2$~\cite{kl3:Nf2:ETM} QCD 
are consistent with our result.


\begin{figure}[b]
\begin{center}
   \includegraphics[angle=0,width=0.48\linewidth,clip]%
                   {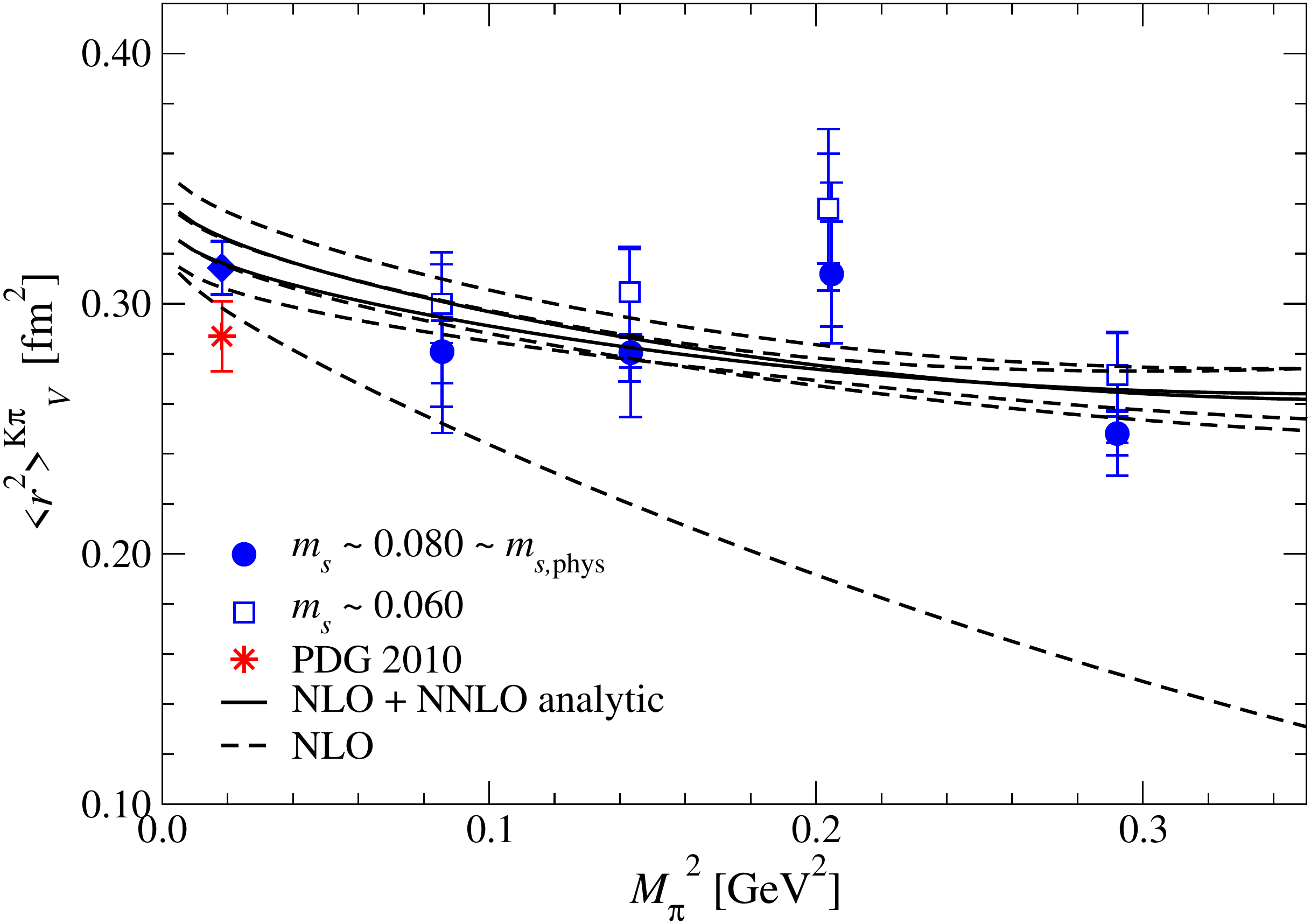}
   \hspace{3mm}
   \includegraphics[angle=0,width=0.48\linewidth,clip]%
                   {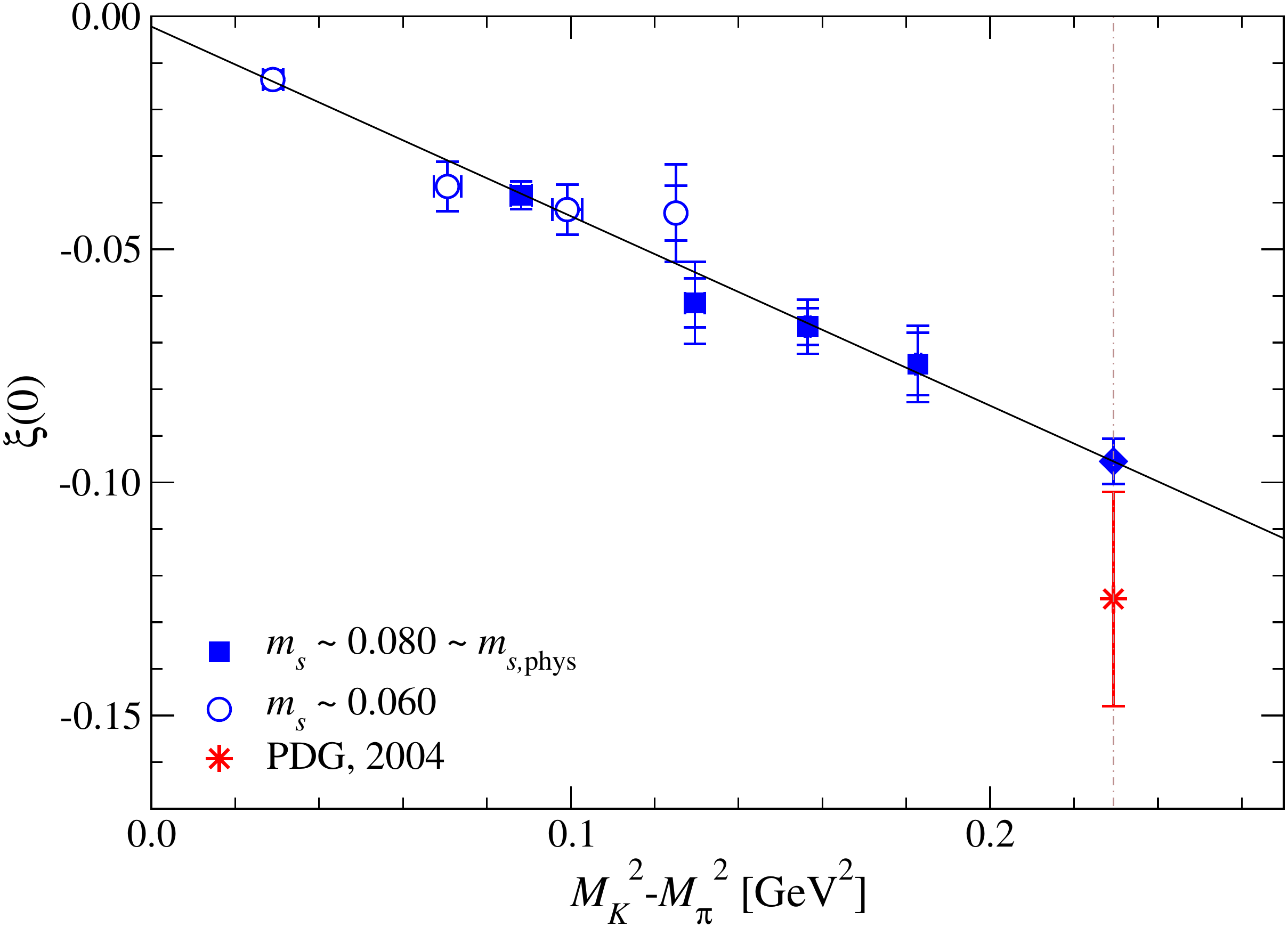}
   \vspace{-2mm}
   \caption{
      Chiral extrapolation of $\crad^{K\pi}_V$ (left panel) and 
      $\xi(0)$ (right panel).
      In both panels,
      the diamonds represent our result extrapolated to the
      physical quark masses $m_{ud}$ and $m_s$,
      which should be compared to the experimental values plotted by the stars.
   }
   \label{fig:dble_rat:r2_xi}
\end{center}
\vspace{-5mm}
\end{figure}

\section{Comparison of $\crad^{K\pi}_V$ and $\xi(0)$ with ChPT and experiments}

In order to demonstrate the reliability of the 1\,\% calculation of $f_+(0)$,
we compare our data of the normalized slope 
$\crad^{K\pi}_V\!=\!6f_+^{\prime}(0)/f_+(0)$ 
and $\xi(0)$ with ChPT and experiments. 
In contrast to $f_+(0)$, 
the Ademollo-Gatto theorem is not applicable to these quantities,
and unknown LECs appear 
already in their leading chiral corrections.
For $\crad^{K\pi}_V$,
we parametrize its higher order contributions
by simple analytic terms
to avoid unstable chiral extrapolations
\bea
   \crad^{K\pi}_V
   =
   12L_9^r/F_\pi^2 + \mbox{``chiral logarithms''} + d_\pi\xi_\pi +¡¡d_K\xi_K.
\eea
We refer to Ref.~\cite{kl3:ChPT:f2} for the explicit expression 
of the chiral logarithmic terms.
The left panel of Fig.~\ref{fig:dble_rat:r2_xi}
shows that 
this form describes our data reasonably well.
It also suggests that the chiral behavior of $\crad^{K\pi}_V$ 
is significantly modified in our simulation region
$290 \!\lesssim\! M_\pi\mbox{[MeV]} \!\lesssim\!540$
by the $O(\xi_P)$ contribution at two-loop order in ChPT.
We note that significant two-loop corrections have been also observed 
in our studies of the pion and kaon charge radii~\cite{prev0,prev1}.
We confirm a good agreement of the extrapolated value of $\crad^{K\pi}_V$
with experiment~\cite{PDG:2010}.
Our fit result
$L_9^r \times 10^3 \!=\! 4.3(0.6)_{\rm stat}(0.3)_{\rm sys}$
is also consistent with the phenomenological estimate 5.9(0.4)~\cite{L9:BT}.

In this report,
we parametrize the quark mass dependence of $\xi(0)$ by a simple linear form
$\xi(0) = d_0 + d_1 (M_K^2-M_\pi^2)$, 
which is motivated from the ChPT expression of the leading analytic terms 
$\propto M_K^2-M_\pi^2$~\cite{kl3:ChPT:f2}.
This form describes our data reasonably well
as seen in the right panel of Fig.~\ref{fig:dble_rat:r2_xi}. 
We obtain $d_0\!=\!-0.022(25)$ confirming that $\xi(0)$ vanishes 
in the $SU(3)$ symmetric limit, as expected.
The extrapolation to the physical point yields 
$\xi(0)\!=\!-0.095(5)$
which is consistent with the experimental value 
$-0.125(23)$~\cite{PDG:2004}.


\section{Summary}
\label{sec:summary}


In this article,
we report on our calculation of the kaon semileptonic form factors.
The normalization $f_+(0)$ is calculated within 1\,\% accuracy
by utilizing the all-to-all quark propagator, reweighting and TBCs.
The reliability of this precise calculation is checked by 
confirming a good consistency of $\crad_V^{K\pi}$ and $\xi(0)$ 
with experimental results.


Since we observe significant two-loop contributions 
in the chiral expansions of $f_+(0)$ and $\crad_V^{K\pi}$,
it is interesting to apply two-loop ChPT formulae to our data.
To this end,
our use of the overlap quark action is advantageous,
since exact chiral symmetry enables us 
to use the two-loop formulae in the continuum limit 
without any additional terms at finite lattice spacings. 
This provides a theoretically clean comparison 
between lattice QCD and ChPT at the higher order.

\vspace{3mm}

Numerical simulations are performed on Hitachi SR11000 and 
IBM System Blue Gene Solution 
at High Energy Accelerator Research Organization (KEK) 
under a support of its Large Scale Simulation Program (No.~11-05)
as well as on Hitachi SR16000 at YITP in Kyoto University.
This work is supported in part by 
the Grants-in-Aid for Scientific Research 
(No.~21674002, 21684013), 
the Grant-in-Aid for Scientific Research on Innovative Areas
(No. 2004: 20105001, 20105002, 20105003, 20105005, 23105710),
and SPIRE (Strategic Program for Innovative Research).

\end{document}